\documentclass[a4paper]{jltp}  

\usepackage{amssymb}
\usepackage{graphicx}

\graphicspath{{Fig/}}

\title{
  What is wrong with paramagnons?
} 

\author{Hartmut Monien}

\address{
  Physikalisches Institut,
  Universit\"at Bonn, 
  53115 Bonn, Germany
}

\begin{document} 

\maketitle

\begin{abstract} 
  Systems with itinerant fermions close to a zero temperature quantum phase
  transition like the high temperature superconductors exhibit unusual
  non-Fermi liquid properties. The interaction of the long-range and
  low-energy fluctuations of the incipient order with the fermions modify the
  dynamical properties of the fermions strongly by inducing effective
  long-range interactions. Close to the transition the interaction of the
  order parameter fluctuations becomes important. In this paper we discuss the
  physics of the non-Gaussian order parameter fluctuations on the electronic
  spectrum and illustrate their effect by considering the charge-density-wave
  transition and a two dimensional superconductor. 
\end{abstract}

\section{INTRODUCTION}
Fermionic systems which undergo a zero temperature phase transition as a
function of some experimental parameter exhibit unusal physical properties.
The phase transition which are driven by quantum rather than thermal
fluctuations have been studied intensively theoretically and experimentally
the last few years.  While significant progress in the understanding of low
dimensional spin systems could be achieved the physics of itinerant electrons
coupled to the low energy collective fluctuations is far less understood.

One of the problems studied most intensively in this context is a
itinerant electron magnetism.  The interplay between superconductivity
and spin fluctuations in nearly ferromagnetic metals has been first
studied in nearly ferromagnetic metals.  Berk and
Schrieffer\cite{BerkSchrieffer66} and
Doniach\cite{DoniachEngelsberg66} used the term "paramagnons" to
describe the long-lived and long-range spin fluctuations in a nearly
ferromagnetic metal.  The exchange of long-lived and long-range
ferromagnetic spin fluctuations has been argued to mediate an unusual
type of superfluidity in $^3$He.  The nonlinear feedback of
spin-fluctuations has been used sucessfully to explain the stability
of the "Anderson-Brinkmann-Morel" superfluid phase.  However by now it
has become clear that $^3$He closer to localization than to a
ferromagnetic transition\cite{VollhardtWoelfle90}.  In recent years it
has been proposed that in the high temperature superconductors with
their magnetic parent compounds the exchange of antiferromagnetic spin
fluctuations may lead to d-wave superconductivity
\cite{Scalapino86,Pines91}. Some experimental features like the
the unusual relaxation rates in the NMR experiments have been
understood using a {\em phenomenologicalical approach} to the spin
fluctuations\cite{MMP90}.  Various re-summation schemes,
like the fluctuation-exchange approximation (FLEX) and other more
complicated schemes like Parquet re-summation\cite{Bickers89}, have
been used to calculate properties in the "nearly antiferromagnetic
phase" and to compare it to more {\em microscopic} approaches. Already
in 1976 Hertz showed how to generalize the concepts of classical phase
transitions to a zero temperature quantum phase transition for
itinerant electrons\cite{Hertz76}. While thermodynamic properties of
the electronic system can be calculated close to the phase transition
the spectral properties of the electrons have been calculated only in
approaches basically relying on the assumption of Gaussian order
parameter fluctuations\cite{Millis93}. Very little is known about the
physics of the electronic degrees of freedom in the non-Gaussian
regime. The purpose of the paper is to draw attention to the fact that
the non-Gaussian order parameter fluctuations drastically modify the
electronic properties.  This is particularly relevant for the {\em
  pseudo-gap regime}.  The paper is organized as fowllows. First we
briefly review the theory of {\em quantum phase transitions} by Hertz
and Millis. We then explain the physics for of the non-Gaussian regime
using a one-dimensional example. Finally we apply how to generalize
the one-dimensional example to the problem of phase fluctuations in a
two-dimensional superconductor.

\section{QUANTUM PHASE TRANSITIONS}
The partition , $Z$, function of an interacting electron gas with incipient
order can be expressed as a functional integral over a field
$\Phi$ which represents the order-parameter density of the incipient
order, and the electronic degrees of freedom, $\psi$\cite{Schulz1990}:
\begin{equation}
  \label{eq:PartitionFunction}
  Z = Z_0 \int{\cal D}[\Phi]  \int {\cal D}[\psi] \exp(-S[\psi, \Phi ])
\end{equation} 
The last integal defines an effective action, $S_{eff}[\Phi]$, for the
bosonic field $\Phi$:
\begin{equation}
  \label{eq:Seff}
  S_{eff}[\Phi] = \int{\cal D}[\psi]\exp(-S[\psi, \Phi ])
\end{equation}
The electronic degrees of freedom can be integrated out, since they
only appear quadratic in the action.  An external field $h_n(q)$ can
be coupled to the field $\Phi$ via
\begin{equation}
  \label{eq:coupling}
  S_{eff}^{(ext)} = \beta\sum_{i\omega_n, q} h_n(q)\cdot\Phi_{-n}(-q)
\end{equation}
Integrating out the electronic degrees of freedom and expanding in the
field $\Phi$ the action $S[\Phi]$ can be written as
\begin{equation}
  \label{eq:SExpansion}
  S_{eff}[\Phi] = S_{eff}^{(0)} 
  + S_{eff}^{(2)}[\Phi] 
  + S_{eff}^{(4)}[\Phi] 
  + \ldots  
  \end{equation}
as shown in Fig. {\ref{fig:expansion}. The zig-zag line represents the 
field $\Phi$ and the solid line the electron propagator.
\begin{figure}[ht]
  \begin{center}
    \includegraphics*[width=\textwidth]{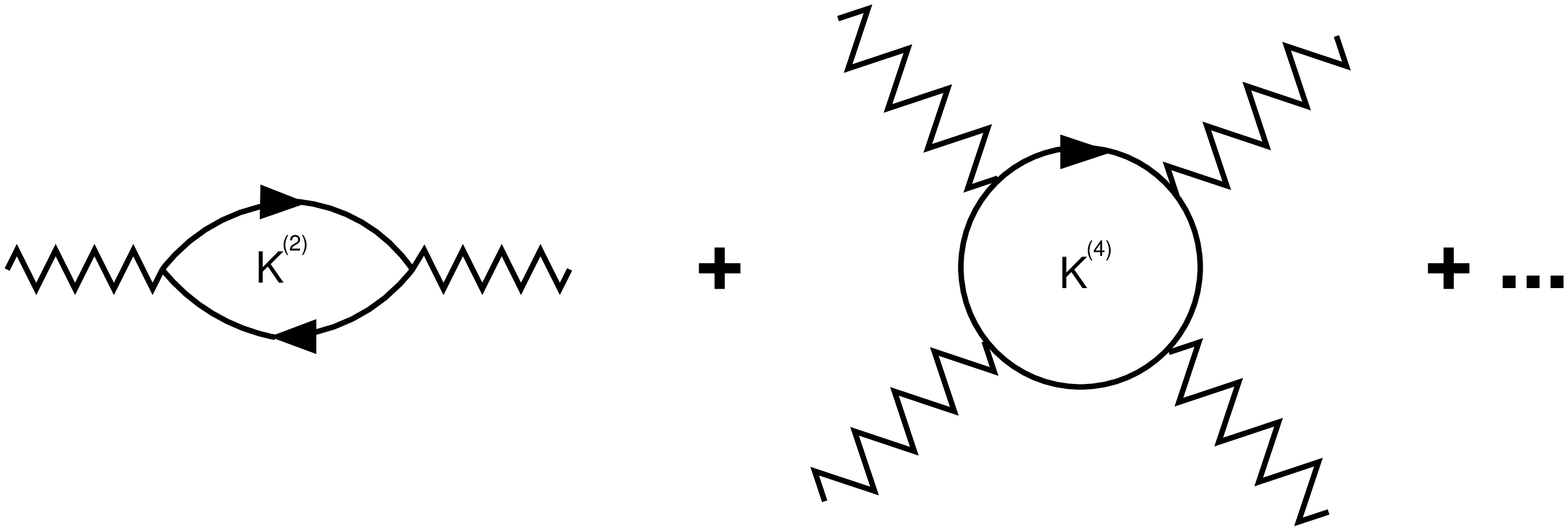}
  \end{center}
  \caption{
    Expansion of the partion function $Z$ in terms of the
    Hubbard-Stratanovich fields.  
    }
  \label{fig:expansion}
\end{figure}

The second order term is given by:
$
  \label{eq:S2}
  S^{(2)}[\Phi] = \beta \sum_{1, 2}
  K^{(2)}(1,2) 
  \Phi(1)\Phi(2)
$
where $1$ represents $(q_1, i\omega_1,\alpha_1)$ and so on. The sum over the
bosonic Matsubara frequencies $i\omega$ is multiplied with the
inverse temperature $\beta$. $K^{(2)}(1,2)$ can be calculated explicitely
$
  K^{(2)}(1,2)  = 
  \delta_{\omega_1+\omega_2}
  \delta_{q_1 + q_2}
  \delta_{\alpha,\alpha'}
  U \left(1-U\chi_0(q_1, i\omega_1)\right)
$
where $\chi_0(q_1, i\omega_1)$ is the electronic susceptibility and 
$U$ is the coupling strength of the electrons to $\Phi$.
\begin{equation}
  \label{chi0}
  \chi_0(q, i\omega) = 
  -\sum_p\frac
  {f(\epsilon_{p+q}) - f(\epsilon_{p})}
  {\epsilon_{p+q}-\epsilon{p}-i\omega}  
\end{equation}
where $f(\epsilon_p)$ is the Fermi distribution function. In a
paramagnon type of theory only $K^{(2)}$ is considered and the 
interaction of the the bosonic fields is neglected. In the
self-consistent renormalized (SCR) approach by Moriya \cite{Moriya85}
the interaction of the bosonic modes is taken into account by
renormalizing $K^{(2)}$ . The SCR therefore an effective gaussian
theory. The effects we are considering here are due to the interaction
of the bosonic modes which is given by the next term in the series for
$S_{eff}$.

The fourth order term obviously depends on four bosonic fields and is given by:
$
  S^{(4)}[\Phi] =  
  \beta\sum_{1,2,3,4}
  K^{(4)}(1, 2, 3, 4)
  \Phi(1)\Phi(2)\Phi(3)\Phi(4)
$.
If the Fermi surface is not nested and the ordering wave vector of the
incipient order is not an extremal spanning vector of the Fermi surface then
the complicated fourth order term might be approximated by a constant. The
effective action including the fourth order term was first studied by Hertz
\cite{Hertz76} and subsequently by Millis\cite{Millis93} using renormalization
group. The resulting phase diagram for a two dimensional system is sketched in
Fig.  \ref{fig:Millis-phasediagram}.   
\begin{figure}[ht]
  \begin{center}
    \includegraphics*[width=0.75\textwidth]{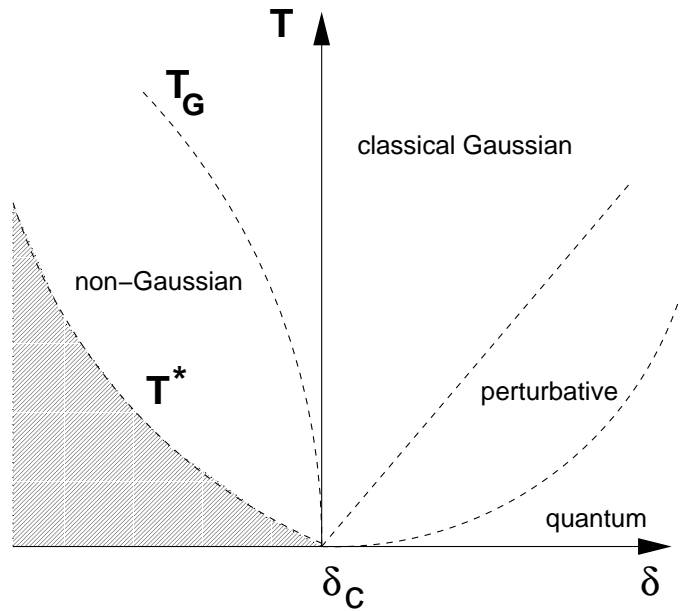}
  \end{center}
  \caption{
    The phase diagram for an itinerant electronic system close to a
    quantum phase transition controled by a parameter $\delta$. The
    dashed lines refer to cross-overs. The region relevant for this
    paper is located to the left of $\delta_c$.
    }
  \label{fig:Millis-phasediagram}
\end{figure}
We briefly summarize the results by Millis relevant for this paper.
To the left of the quantum critical point the phase diagram is
dominated by three regions. On lowering the temperature one first
enters a regime which is dominated by classical order parameter
fluctuations. Below the doted line which indicates the Ginzburg
temperature classical non-Gaussian order parameter fluctuations occur.
Decreasing the temperature further the behavior on the model depends
on the symmetry class of the fluctuations and whether the fermions are
fully or partly gapped.  Paramagnon like theories are often used in a
parameter regime close to a 3D transition which is intercepting the
low temperature regime of the quantum phase transition. It is often
assumed that the order parameter fluctuations are still gaussian.
However in a regime where the electronic density of states is already
substantially suppressed like in the pseudogap regime this is a
questionable assumption.  The purpose of this paper is draw attention
to the fact that the non-Gaussian fluctuations modify the electronic
properties drastically compared to the Gaussian fluctuations.

\section{A QUASI-ONE-DIMENSIONAL PROBLEM}
To illustrate the effect of the non-gaussian fluctuations we discuss a
model system for which most properties can be calculated without
further approximation.  Contrary to the assumption in previous work
\cite{LeeRiceAnderson73,Sadovski79}, it is {\em not} sufficient to
describe the order parameter fluctuations by the {\em variance} and
{\em correlation length} only, but that one needs to consider higher
moments of the order parameter correlator.  This becomes intuitively
clear if one considers the situation depicted in Fig.
\ref{fig:non-gauss}.
\begin{figure}[ht]
  \begin{center}
    \includegraphics*[width=0.75\textwidth]{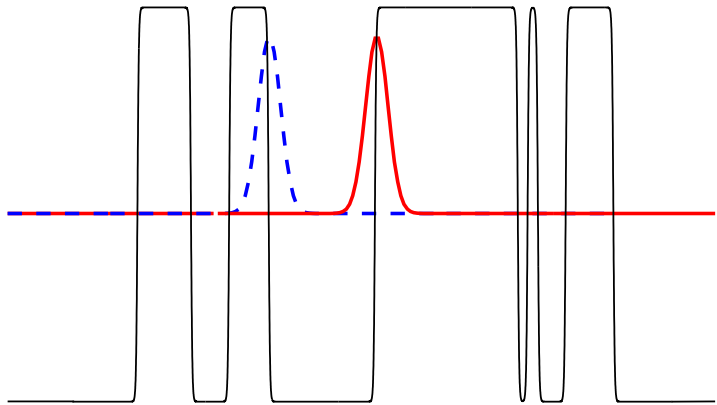}
  \end{center}
  \caption{
    Typical order parameter configuration in the 
    non-Gaussian regime. The electronic wave function is 
    localized in the regions where the order-parameter
    is small. The overlap between electronic wave function
    localized in different regions is exponentially small.
    }
  \label{fig:non-gauss}
\end{figure}
If the order parameter varies smoothly, as in the Gaussian regime,
regions where the order parameter is suppressed are smeared out over
the correlation length.  The electronic wavefunction is spread out
over a length comparable with the correlation length. The kinetic
energy is low and consequently many states can be found at low energy
even when the correlation length is large. On the other hand, if the
order parameter is established and only suppressed over a length scale
much shorter than the correlation length of the potential, as in the
non-Gaussian regime, the electronic wavefunction decays over a
distance $v_F/\Delta$ and has a large kinetic energy.  For the {\em
  same} correlation length and variance of the order parameter the
electronic wavefunction is much stronger suppressed for non-Gaussian
fluctuations. The overlap of parts of the wavefunction located at
regions where the order parameter is exponentially small. In dimension
$d$ the electron wave function will be constrained to a $d-1$
dimensional manifold. If the dimension is greater than one the regions
where the order parameter is small form a ``network'' in which the
electrons can move and one has to consider ``quantum-percolation''.

To be specific we consider here only the most simple case namely a
one-dimensional system close to a CDW transition. The dispersion of
the electrons close to the Fermi energy can be assumed to be linear.
The Hamiltonian of the electrons has the form:
\begin{equation}
  \label{eq:Hamiltonian}
  \hat H = 
  -i v_F (R^\dagger \partial_x R - L^\dagger \partial_x L)
  + \Delta(x)   R^\dagger L
  + \Delta^*(x) L^\dagger R,
\end{equation}
where the operators $R^\dagger$ and $L^\dagger$ create left and right
moving electrons respectively.  The classical order parameter field
$\Delta(x)$ is determined by a Ginzburg Landau action given below,
$v_F$ is the Fermi velocity.  

Next we consider the order parameter fluctuations. For commensurate
fluctuations, the low energy electronic density of states is dominated
by the Dyson singularity which only exists in one
dimension~\cite{MillisMonien00,BartoschKopietzPRB00}. For the more
general case, the order parameter fluctuations are complex and the
Dyson singularity is absent. Therefore we will restrict our discussion
to complex order parameters. After integrating out the electrons the
{\em classical} complex order parameter fluctuations, $\Delta(x)$ are
described by the Ginzburg-Landau action:
\begin{equation}
  \label{eq:order_parameter_action}
  F[\Delta(x)] = \frac{1}{k_B T}\int_0^L\;dx/\xi_0
  \left( 
    \xi_0^2 |\partial_x \Delta|^2 
    + \alpha |\Delta|^2 
    + \beta  |\Delta|^4
  \right)
  \label{eq:F}
\end{equation}
Close to the mean field phase transition $\alpha$ varies linearly with
the temperature $\alpha = \alpha' (T/T_c^{MF}-1)$, whereas $\beta$ and
$\xi_0$ are nearly temperature independent. In principle, the
coefficients $\xi_0$, $\alpha$ and $\beta$ have to be determined self
consistently from the electronic properties. The correlator for the
order parameter fluctuations always decays exponentially,
$<\Delta(x)\Delta(0)> = <\Delta^2> \exp(-|x|/\xi)$ since the 1D system
is disordered above the 3D ordering temperature, $T_c^{3D}$.
Nevertheless the action Eq.~(\ref{eq:order_parameter_action}) has two
different regimes: if $\alpha(T)$ is positive and large, the order
parameter fluctuations are centered around zero and basically
Gaussian. For $\alpha(T)$ negative and large the amplitude of the
order parameter is given by $\sqrt{<\Delta^2>}$ and only the phase
fluctuations play a role. 

The density of states of the electrons in an order parameter potential
given by Eq. (\ref{eq:F}) can be calculated exactly numerically using
the tranfer-matrix formalism for the order-parameter
fluctuations\cite{Monien01}. The Schr\"odinger equation for the
electrons right moving and left moving electrons, with the wavefunction
$u(x)$ and $v(x)$ respectively, reads:
\begin{eqnarray}
  -i v_F \partial_x u(x) + \Delta{\phantom{^*}}(x) v(x) &=& \epsilon u(x)\\
  +i v_F \partial_x v(x) + \Delta^*(x) u(x) &=& \epsilon u(x)
  \label{eq:uv}
\end{eqnarray}
where $\epsilon$ is the single-particle energy.  
The logarithms of $u(x)$ and $v(x)$ obey the following equation of motion:
\begin{eqnarray*}
  -i v_F \partial_x \log(u(x)) &=& \epsilon - 
  \Delta\phantom{^*}(x) u(x)/v(x)\\  
  +i v_F \partial_x \log(v(x)) &=& \epsilon - 
  \Delta^{*}(x) v(x)/u(x)
\end{eqnarray*}
Where one has to keep in mind choosing the correct branch for
$\log(z)$.  Adding the two equation we obtain an equation for the
ratio $r(x) = u(x)/v(x)$:
\begin{equation}
  \label{eq:ratio}
  -i v_F \log(r(x)) = 2
  \left[
    \epsilon - \Delta(x) r(x) - \Delta^*(x) r^{-1}(x)
  \right]
\end{equation}
With the ansatz $r(x) = \exp(i\varphi(x))$ and $\Delta(x) =
|\Delta(x)|e^{i\Phi}$ we finally obain the equation of motion 
for the $\varphi(x)$:
\begin{equation}
  \label{eq:phi}
  \partial_x\varphi(x) = 2
  \left[
    \epsilon - |\Delta(x)|\cos(\varphi(x)+\Phi(x))
  \right]
\end{equation}
Note that $\varphi(x)$ is the scattering phase of the wavefunction.
Surprisingly the phase information alone is quite sufficient to obtain
all the thermodynamic information.  The integrated density of states,
$\cal{N}(\epsilon)$, can be related to the phase by\cite{Lifshits88}
\begin{equation}
  \label{eq:integrated-dos}
  {\cal{N}}(\epsilon) = \lim_{L\rightarrow\infty}\frac{\varphi(L)}{2L}
\end{equation}
The density of states can then be obtained by numerically
differentiating ${\cal{N}}(\epsilon)$ with respect to $\epsilon$. 
\begin{figure}[ht]
  \begin{center}
    \includegraphics*[width=0.9\textwidth]{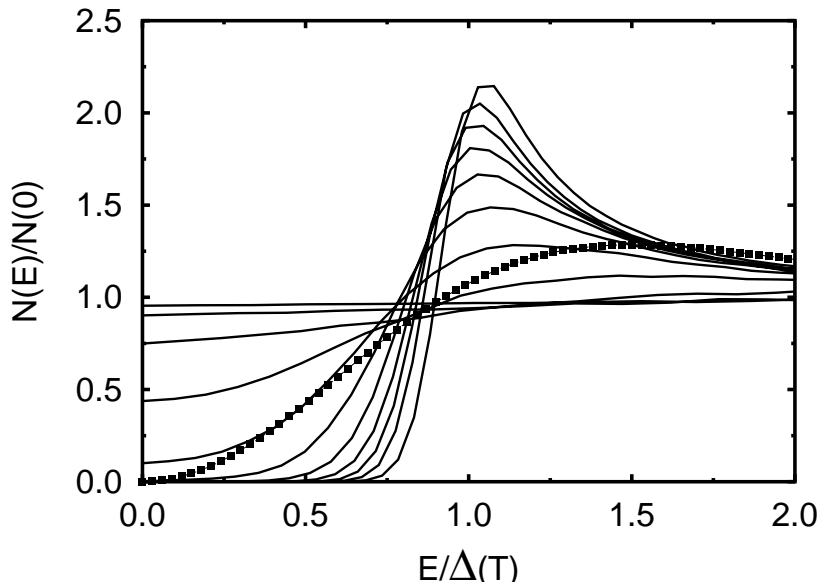}
  \end{center}
  \caption{
    The electronic density of states as a function of temperature.
    The temperatures are 0.1, 0.2, 0.3 ... 1.0 $\times\;T^{MF}_c$.
    The Ginzburg temperature is set to $T_G = 0.1 T^{MF}_c$.
    The square boxes are the result of the Gaussian calculation
    in the limit correlation length going to infinity.
    }
  \label{fig:dos-1D}
\end{figure}
Calculating $\varphi(x)$ for a particular configuration of the the
order-parameter field amounts to integrating a one-dimensional
differential equation with a random potential $\Delta(x)$. 

The remaining problem is to generate a typical configuration of
$\Delta(x)$ and sum over all configurations with the correct
statistical weight. In principle this can be done using a classical
Monte-Carlo simulation. It turns out that there is much more efficent
way for this problem since eigenfunctions of the the transfer matrix
for the free energy function Eq. \ref{eq:F} can be calculated easily.
Using the transfer matrix formalism this problem can be mapped to a
Langevin equation for $\Delta(x)$:
\begin{equation}
  \label{eq:Langevin}
  \frac{\partial\Delta}{\partial x} = \exp(-i\Phi) f(|\Delta|) + \eta
\end{equation}
where $f$ is the ``guiding functions'' of the random walk and $\eta$
is complex Gaussian white noise.  If $f(|\Delta|)$ is a linear function
then Eq. \ref{eq:Langevin} just describes an Orenstein-Uhlenbeck
process with a Gaussian spatial correlation. It has been shown that
$f$ can be obtained from the transfer-matrix of the Eq. \ref{eq:F}.

The exact result for the density of states of the electrons as a
function of temperature is shown in Fig. \ref{fig:dos-1D} and compared
with the result of the calculation assuming Gaussian fluctuations with
a very long correlation length \cite{Sadovski79}. Whereas the Gaussian
model still allows many states in the mean-field gap the non-Gaussian
fluctuations remove the states much more effectively. The artefact in
the Gaussian model that the density of states behaves like
$N(\epsilon)\sim\epsilon^2$ even when the system is ordered is due to
the fact that in the Gaussian model the most probable value for the
order parameter is $|\Delta|=0$ which is not true in the non-Gaussian
model. In the Gaussian theories the surpression of the electronic
states is due to large amplitude of the order parameter fluctuations
which are strongly overestimated if one would like to obtain the same
loss of spectral weight as in the non-Gaussian model.  Since the
paramagnon theories are Gaussian the naive application to the
pseudogap regime is highly questionable\cite{Schmalian98,Schmalian99}.

\section{QUASIPARTICLES IN A TWO-DIMENSIONAL SUPERCONDUCTOR}

In this section we apply the methods developed in the previous chapter
to the thermodynamic properties of a quasi-particle in a
two-dimensional superconductor. It has been claimed that the thermal
and transport properties of a high temperature superconductor above
the transition are mostly determined by the phase-fluctuations of the
preformed superconducting pairs. We will apply the quasi-classical
approximation which is not really justified if the length scale of the
phase fluctuations of the order parameter is comparable to the lattice
spacing. But in that case we will recover the ``normal'' quasi-particles
anyways.

It has been noted by Schopohl\cite{Schopohl98} that the quasi-classical
Eilenberger equation of a superconductor can be mapped to the
linearized Bogoliubov-DeGennes (BdG) equation. If we identify $u$ and
$v$ with particle and hole excitation in a superconductor Eq.
\ref{eq:uv} can be viewed as a BdG equation. With the phase equation,
Eq. \ref{eq:phi} it then is feasible to calculate the density of
states of an inhomogeneus superconductor relatively easily.
\begin{figure}[ht]
  \begin{center}
    \includegraphics*[width=0.8\textwidth]{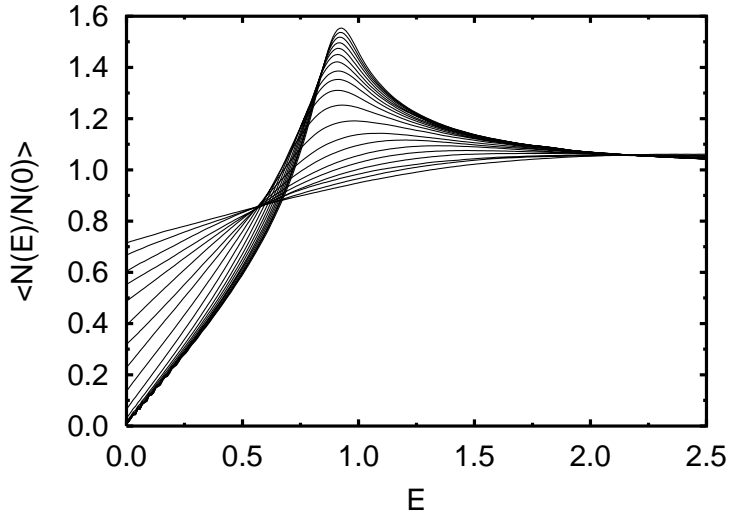}
  \end{center}
  \caption{
    Density of states of a quasiparticle in a two-dimensional
    $d_{x^2-y^2}$ superconductor. The inverse temperature $\beta =
    1/T$ is varied from 0.1 to 2.0 in steps of 0.1. The BKT transition
    is located approximately at $J \beta_c\approx0.7$, where $J$ is
    the coupling of the phases.  The energy is measured in terms of
    the amplitude of the order parameter.  
    }
  \label{fig:dos-2D}
\end{figure}
We have applied this method to calculate the density of states of a
quasiparticle put into a two-dimensional superconductor. We assume
that the order parameter amplitude is already fixed. The phase
fluctuations are then described by an xy-model. The classical xy-model
can be simulated efficiently even close to the
Bereshinskii-Kosterlitz-Thouless (BKT) transition using the Wolff
algorithm\cite{Wolff89}. We used a 128 x 128 grid for the xy - model.
For each thermal configuration the phase equation is solved and the
density of states is calculated. Finally we assume a $d_{x^2-y^2}$ gap
- which is the relevant gap structure for the high temperature
superconductors. The result of the calculation is shown in Fig.
\ref{fig:dos-2D}. The temperature is varies from much below the BKT
transition to very high temperatures. The suprising fact is that the
BKT transition has very little effect on the equilibrium electronic
degrees of freedom. At very high temperatures on does recover most of
the density of states although the order parameter is already
established. At very low temperatures one recovers the usual d-wave
density of states for a $d_{x^2-y^2}$ superconductor, which is linear
at low energies\cite{Hirschfeld86,Monien87}.  It would be
interesting to apply the quasi-classical approach to the transport
properties in the phase-fluctuation regime.  However for transport 
vortices have a highly nontrivial effect on the electrons.

\section{CONCLUSIONS}
Paramagnon like theories have been used extensively to study
electronic properties of itinerant electronic systems close to a phase
transition. These theories amount to using a Gaussian approximation
for the incipient order parameter field.  In this paper we have
discussed the effect of non-Gaussian order parameter fluctuations on
the electronic properties. The non-Gaussian order parameter
fluctuations have been shown to be relevant for the understanding of a
two-dimensional itinerant system close to a quantum phase transition.
For a one-dimensional system close to a charge-density-wave transition
the electronic properties can be calculated exactly. The suppression
of the density of states is much more dramatic than in Gaussian
models.  Finally we have investigated a
quasiparticle in a two-dimensional superconductor and mapped the
problem to a one-dimensional BdG equation.

{\em Dedication:} This paper is dedicated to Peter W\"olfle on the 
occasion of his 60th birthday.}
\bibliography{qpt}

\begin{thebibliography}{10}

\bibitem{BartoschKopietzPRB00}
L.~Bartosch and P.~Kopietz.
\newblock Classical phase fluctuations in incommensurate peierls chains.
\newblock {\em Phys. Rev. {\bf B}}, 62(24):{\bf R} 16223, 2000.

\bibitem{BerkSchrieffer66}
N.~F. Berk and J.~R. Schrieffer.
\newblock Effect of {F}erromagnetic {S}pin {C}orrelations on
  {S}uperconductivity.
\newblock {\em Phys. Rev. Lett.}, 17:433, 1966.

\bibitem{Bickers89}
N.~E. Bickers and D.~J. Scalapino.
\newblock Conserving approximation for strongly fluctuating electron systems 1.
  formalism and calculational approach.
\newblock {\em Ann. Phys. (N.Y.)}, 193:206, 1989.

\bibitem{DoniachEngelsberg66}
S.~Doniach and S.~Engelsberg.
\newblock Low-{T}emperature {P}roperties of {N}early {F}erromagnetic {F}ermi
  {L}iquids.
\newblock {\em Phys. Rev. Lett.}, 17:750, 1966.

\bibitem{Hertz76}
J.~Hertz.
\newblock Quantum critical phenomena.
\newblock {\em Phys. Rev. {\bf B}}, 14:1165, 1976.

\bibitem{Hirschfeld86}
P.~Hirschfeld, D.~Vollhardt, and P.~W\"olfle.
\newblock {\em Solid State Commun.}, 59:111, 1986.

\bibitem{LeeRiceAnderson73}
P.~A. Lee, T.~M. Rice, and P.~W. Anderson.
\newblock Fluctuation effects at a peierls transition.
\newblock {\em Phys. Rev. Lett.}, 31(7):462--465, 1973.

\bibitem{Lifshits88}
I.~Lifshits and L.~A.~Pastur S.~A.~Gredeskul.
\newblock {\em Introduction to the theory of disordered systems}.
\newblock John Wiley \& Sons, N. Y., 1988.

\bibitem{Millis93}
A.~J. Millis.
\newblock Effect of a nonzero temperature on quantum critical points in
  itinerant fermion systems.
\newblock {\em Phys. Rev. {\bf B}}, 48:7183, 1993.

\bibitem{MillisMonien00}
A.~J. Millis and H.~Monien.
\newblock On pseudogaps in one-dimensional models with quasi-long-range-order.
\newblock {\em Phys. Rev. {\bf B}}, 61:12496, 2000.

\bibitem{MMP90}
A.~J. Millis, H.~Monien, and D.~Pines.
\newblock Phenomenological model of nuclear relaxation in the normal state of
  {YB}a2{C}u3{O}7.
\newblock {\em prb}, 42(1):167--178, 1990.

\bibitem{Monien01}
Hartmut Monien.
\newblock Exact results for the {C}rossover from {G}aussian to {N}on-{G}aussian
  order {P}arameter {F}luctuations in {Q}uasi-{O}ne-{D}imensional {E}lectronic
  {S}ystems.
\newblock {\em Phys. Rev. Lett.}, 87:126402, 2001.

\bibitem{Monien87}
{K. and Walker D.} Monien, {H. and Scharnberg}.
\newblock Resonant {I}mpurity {S}cattering in {A}nisotropic {S}uperconductors:
  {E}ffects of {A}rbitrary {P}hase {S}hifts and {P}article {H}ole {A}symmetry.
\newblock {\em Solid State Commun.}, 63:263, 1987.

\bibitem{Moriya85}
T.~Moriya.
\newblock {\em Spin fluctuations in itinerant electron magnetism}, volume~56.
\newblock Springer series in solid-state sciences, Berlin, New York, 1985.

\bibitem{Schopohl98}
{N. Schopohl}.
\newblock Transformation of the {E}ilenberger {E}quations of
  {S}uperconductivity to a {S}calar {R}iccati {E}quation.
\newblock (cond-mat/9804064), 1998.

\bibitem{Pines91}
D.~Pines.
\newblock Spin fluctuations and high temperature superconductivity in the
  antiferromagnetically correlated oxides ybco and lsco.
\newblock {\em Physica {\bf C}}, 185:120--129, 1991.

\bibitem{Sadovski79}
M.~V. Sadovskii.
\newblock Exact solution for the density of electronic states in a model of
  disordered system.
\newblock {\em Sov. Phys. JETP}, 50:989, 1979.
\newblock Zh. Eksp. Teor. Fiz. {\bf 77}, 2070 (1979).

\bibitem{Scalapino86}
D.~J. Scalapino, Jr. E.~Loh, and J.~E. Hirsch.
\newblock {d}-wave pairing near a spin-density-wave instability.
\newblock {\em Phys. Rev. {\bf B}}, 34(11):{\bf R}8190--8192, 1986.

\bibitem{Schmalian98}
{D and Stojkovic} Schmalian, {J and Pines}.
\newblock Weak pseudogap behavior in the underdoped cuprate superconductors.
\newblock {\em prl}, 80(17):3839--3842, 1998.

\bibitem{Schmalian99}
{D and Stojkovic} Schmalian, {J and Pines}.
\newblock Microscopic theory of weak pseudogap behavior in the underdoped
  cuprate superconductors: {G}eneral theory and quasiparticle properties.
\newblock {\em prb}, 60:667--686, 1999.

\bibitem{Schulz1990}
H.~J. Schulz.
\newblock Effective {A}ction for strongly correlated fermions from functional
  integrals.
\newblock {\em Phys. Rev. Lett.}, 65(19):2462--2465, 1990.

\bibitem{VollhardtWoelfle90}
D.~Vollhardt and P.~W\"olfle.
\newblock {\em The superfluid phases of $^3$He}.
\newblock Francis and Taylor, London, 1990.

\bibitem{Wolff89}
U.~Wolff.
\newblock Collective monte carlo updating for spin systems.
\newblock {\em Phys. Rev. Lett.}, 62(4):361, 1989.

\end{thebibliography}

\end{document}